\documentstyle[aps,prb,multicol,epsf,amssymb]{revtex}
\def\gs{\gtrsim}
\def\ls{\lesssim}
\def\be{\begin{equation}}
\def\en{\end{equation}}                  
\def\p{\partial} 
\newcommand{\bi}[1]{\mbox{\boldmath$#1$}}
\def\bea{\begin{equation}\begin{array}{rcl}}
\def\ena{\end{array}\end{equation}}

\def\ge{> \kern -12pt \lower 5pt \hbox{$\displaystyle =$}}
\def\le{< \kern -12pt \lower 5pt \hbox{$\displaystyle =$}}
\def\gs{> \kern -12pt \lower 5pt \hbox{$\displaystyle{\sim}$}}
\def\ls{< \kern -12pt \lower 5pt \hbox{$\displaystyle{\sim}$}}
\def\ep{\epsilon}

\def\be{\begin{equation}}

\def\bea{\begin{eqnarray}}
\def\en{\end{equation}}
\def\ena{\end{eqnarray}}
\def\p{\partial }
\def\ve{\varepsilon} 
\renewcommand{\theequation}{\arabic{section}.\arabic{equation}}

\begin{document}
\draft
\bibliographystyle{prsty}
\title{Ion-induced nucleation in 
polar one-component fluids}
\author{Hikaru Kitamura and Akira Onuki}
\address{Department of Physics, Kyoto University, Kyoto 606-8502, Japan}
\maketitle

\begin{abstract}
{We present a Ginzburg-Landau 
theory   of  ion-induced nucleation in 
 a gas phase of polar 
one-component fluids, where a liquid droplet grows 
with an ion at its center. By calculating  
the density profile around an ion,  we  show that 
the solvation free energy is larger in  gas than in  liquid 
at the same temperature  on the coexistence curve. 
This  difference  much reduces the nucleation 
barrier in a metastable gas.  
}
\end{abstract}

\begin{multicols}{2}

\section{Introduction}

Ion-induced nucleation in water vapor  
plays a decisive role in many atmospheric phenomena. 
It is well-known that   the nucleation rate 
is much enhanced in polar fluids in the presence of ions 
\cite{Wilson,Thomson}. 
However, precise experiments on this effect 
have been difficult despite 
the long history of this problem 
since Wilson's cloud chamber 
experiment \cite{Wilson,Seto,Mora}.   
Theoretically, we need to understand statistical 
behavior of polar molecules  
around  an ion \cite{Is} and 
heterogeneous nucleation triggered by 
such ion-dipole interaction in  metastable states.
This problem is thus of fundamental importance 
in physics and chemistry, 
but it has  rarely been studied in the literature.

Originally, Thomson \cite{Thomson}
 calculated the chemical potential 
of vapor in the vicinity of a charged particle 
using  continuum electrostatics and 
predicted  formation of a small liquid 
droplet with the ion at its center.  However, its radius 
is only $3{\rm \AA}$ in room-temperature water vapor, 
 so Thomson's picture is an oversimplified one.  
Recently,  ion-induced nucleation has been 
studied using a density functional theory 
\cite{Kusaka,Kusaka1}, 
a Monte Carlo method \cite{Zeng}, and molecular dynamics 
simulations \cite{Elena}. 
These numerical studies 
 have shown  clustering of  polar 
molecules around an ion and alignment of  the dipoles 
along the electric field. 
After the statistical average, the first effect gives rise to 
a density inhomogeneity  $n(r)$ 
 or electrostriction around an ion. 
It is worth noting that Born calculated the solvation 
free energy due to an ion in  continuum electrostatics,   
accounting for  polarization alignment (in the linear 
response) but neglecting 
electrostriction \cite{Is,Born,comment}.

The aim of this work is  
to present a Ginzburg-Landau 
 theory of ion-induced nucleation in polar one-component fluids 
in a gas phase.  It will be based 
on our recent theory of solvation effects 
in near-critical polar binary mixtures \cite{NATO,OK}.  Our theory 
will take account of both electrostriction and polarization 
alignment. As  will be discussed, 
the Born formula becomes a poor 
approximation in strong electrostriction, 
particularly when an ion is in a gaseous polar fluids.  
Merits of our approach are  its simplicity and 
its ability to describe mesoscopic 
effects such as nucleation. 
In Section 2 a  theoretical background of solvation 
will be presented.  
In Section 3  the free  energy 
increase to  create a critical droplet around an ion 
will be calculated in a  gaseous polar fluid.

\section{Theoretical Background } 
\setcounter{equation}{0}

\subsection{Electrostatics}

We  place a charged particle (ion) 
in  a   polar one-component  fluid.  
We assume that the static dielectric constant 
$\ve=\ve(n)$ is an increasing 
function of  the fluid density $n$ at each temperature $T$. 
In the continuum approximation the electric field 
 ${\bi E}=-\nabla\Phi$ is 
  induced by the electric charge and 
the electric potential $\Phi$ 
satisfies 
\be 
\nabla\cdot\ve\nabla \Phi=
 -4\pi \rho({\bi r}),  
\label{eq:2.2}
\en  
where $\rho({\bi r})$ is the charge density. 
Here the polarization $\bi P$ 
is assumed to be aligned along $\bi E$ 
as ${\bi P}=\chi{\bi E}$ with 
$\chi=(4\pi)^{-1}(\ve-1)$ being the polarizability 
\cite{comment}. 
 We take the origin of the 
reference frame at the center of the ion. 
It is convenient to 
assume that the charge density is homogeneous within 
a  sphere $r<R_{\rm i}$ as 
\be 
\rho = Ze /( 4\pi R_{\rm i}^3/3)  \quad (r<R_{\rm i}) 
\label{eq:2.8}
\en  
and vanishes outside the sphere $r>R_{\rm i}$.  
The total charge is given by $Ze$. 
In our theory $R_{\rm i}$ is a 
phenomenological parameter representing 
the ion radius \cite{RMarcus}. 
Then $n=n(r)$  and $\Phi= \Phi(r)$ depend 
only on $r=|{\bi r}|$. 
In this case Eq.(2.1) is solved to give 
\be 
-\ve (n) \Phi' (r) 
= \frac{4\pi}{r^2} \int_0^r   dr_1 r_1^2 \rho(r_1)
= \frac{Ze}{r^2}\theta(r),  
\label{eq:2.10}
\en 
where $\Phi'(r)= d\Phi(r)/dr$,  $\theta(r)=1$ for $r>R_{\rm i}$,  and  
\be 
\theta(r)= (r/R_{\rm i})^3\quad (r<R_{\rm i}).
\label{eq:2.11}
\en  
The electrostatic free energy  
is simply  of the form \cite{comment}, 
\be 
F_e= \int d{\bi r} \frac{1}{8\pi}\ve {\bi E}^2
= \frac{1}{2} Z^2e^2 
\int_0^\infty  d{r} \frac{1}{\ve r^2}\theta(r)^2 .
\label{eq:2.12}
\en 
Its functional derivative with respect to $n$ yields 
the chemical potential contribution 
 arising from the electrostatic 
interaction, 
\be 
\mu_e= 
\frac{\delta }{\delta n}F_e= - 
\frac{Z^2e^2\ve'  }{8\pi \ve^2 }  
\frac{\theta(r)^2}{r^4} ,
\en 
where $\ve'= (\p \ve/\p n)_T$. 
This quantity  is negative and 
grows strongly as $r$ approaches  $R_{\rm i}$, 
leading to accumulation of the  
fluid particles around the ion. 
Here we are neglecting  the nonlinear electric field 
effect near the ion \cite{Kusaka1,Abraham,nonlinear}, 
for which see a discussion in the summary. 
The effect arising from the complex molecular 
structure of solvent molecules is also beyond 
the scope of this work,  
which leads to the dependence of the nucleation rate 
on the sign of the charge of ions\cite{Kusaka1}.

A number of theories 
have been presented to describe  
the  overall density and temperature 
dependence of the dielectric constant 
 $\ve= \ve(n,T)$. Recently,  
Fernandez {\it et al.} \cite{Fernandez} analyzed a wide range of 
 data of $\ve$ for water and steam using 
 the formula proposed by Harris and Alder \cite{Alder} 
for polarizable polar fluids, 
\be 
\frac{\ve-1}{\ve+2}= 4\pi n \bigg [ 
 \frac{\alpha_m}{3}   + \frac{g \mu^2}{k_{\rm B}T} 
\frac{\ve}{(2\ve +1)(\ve+2)}\bigg ] ,
\en 
where $\alpha_m$ represents the molecular polarizability and 
$\mu$ is the dipole moment of a molecule($\sim 2$D for water). 
This form reduces to the 
Clausius-Mossotti formula 
in the absence of a permanent dipole  
and tends to the 
Kirkwood formula $(\ve-1)(2\ve +1)/\ve 
= 4\pi n g\mu^2/k_{\rm B}T$ for nonpolarizable 
molecules with permanent dipole moment \cite{Kirkwood}. 
Here $g$ is the so-called Kirkwood correlation factor 
arising from the correlation between the molecular orientations 
due to nondipolar interactions (2$-$3 for  liquid water), 
so $g$ weakly depends on $n$  as revealed 
in  experiments \cite{Archer}. 
In Fig.1 we plot 
$\ve$ as a function of $n$ 
at $T/T_c= 0.675$ ($T=437$ K) 
 and 0.928 ($T=600$ K) below $T_c$ 
on the basis of the formula 
(2.7) using data of  $\alpha_m$ and $g\mu^2$ for water 
in Ref.17.
In our numerical analysis  the quantity 
$\ve'/\ve^2$ in Eq.(2.6) will 
be calculated using the  formula (2.7).

\subsection{Ginzburg-Landau free  energy }

For simplicity we use the  Helmholtz free energy density $f=f(n,T)$ 
in the van der Waals theory \cite{Onukibook}, 
\be
  f =  {k_{\rm B}T}n \ln \bigg [\frac{\lambda_{\rm th}^3n}{1-v_0n}\bigg ]  
    - k_{\rm B}T n-  {4\epsilon_{\rm vw}}{v_0}n^2 , 
  \label{eq:3.4.9} 
   \en
where $v_0$ and $\epsilon_{\rm vw}$ 
represent the molecular hardcore 
volume   and the strength of the 
 attractive part of the pair interaction, respectively, 
and $\lambda_{\rm th}= \hbar (2\pi/mk_{\rm B}T)^{1/2}$ 
is the thermal de Broglie length.  
It follows  the well-known expression for the 
van der Waals pressure,  
\be 
p=  n\mu-f= \frac{k_{\rm B}Tn}{1-v_0n}- 4\epsilon_{\rm vw} v_0n^2 ,
\en 
where $\mu= {\p f}/{\p n}$ is the 
 chemical potential per particle 
 of the form,  
\be
\mu =   
 k_{\rm B}T \ln \bigg [ \frac{\lambda_{th}^3n}{1-v_0n} \bigg ] + 
\frac{k_{\rm B}Tv_0n}{1-v_0n} 
-8\epsilon_{\rm vw} v_0 n.
\label{eq:4.4.1}
\en    
The  isothermal compressibility $K_T$ behaves as 
\be  
{nk_{\rm B} K_T} 
= {(1-v_0n)^2}/[T-T_{\rm s}(n)]  ,  
\en 
where $T_{\rm s}(n)$ is the spinodal temperature given by 
\be 
T_{\rm s}(n)= 8 \epsilon_{\rm vw} v_0n (1-v_0n)^2/k_{\rm B}
\en  
Maximization  of $T_{\rm s}(n)$ with respect to $n$  gives the  
critical temperature and density,  
\be 
T_c=32\epsilon_{\rm vw}/27k_{\rm B}, \quad n_c=1/3v_0. 
\en 
In Fig.2 we show the phase diagram in the $T$-$\phi$ plane 
for the van der Waals fluid, 
where $\phi=v_0n$ is the normalized 
density or the volume fraction of the hard-core region. 
In this work we will set 
$ v_0 = 30.96 \times 10^{-24}$ cm$^3$  and 
$\epsilon_{\rm vw}/k_{\rm B} = 545.98$ K, 
which follow from the 
critical values,  $n_c = 1.076\times 
10^{22}$cm$^{-3}$ and $T_c = 647.096$ K, 
for water \cite{Anisimov,Kiselev}.
We  introduce the van der Waals radius,  
\be 
\sigma= v_0^{1/3}, 
\en 
which is $3.1{\rm \AA}$ for water.

We assume that 
$n$ tends to a homogeneous value 
$n_\infty$ far from the ion $r \rightarrow \infty$ 
 and  that the temperature $T$ is 
homogeneous throughout the system.
These assumptions are allowable in calculating 
the equilibrium solvation profile 
and the critical droplet in nucleation \cite{however}. 
The chemical potential and the pressure   far from the ion 
are written as $\mu_\infty$ and  $p_\infty$, respectively. 
We consider  the grand potential 
$\Omega$ (equal to $-p V$ 
for homogeneous states)  
at the temperature $T$ and the chemical 
potential $\mu=\mu_\infty$.  
Its increase due to an ion including the gradient and electrostatic 
contributions is written as   
\be 
\Delta \Omega = \int d{\bi r}\bigg 
[ g+ \frac{C}{2}|\nabla n|^2 
 + \frac{\epsilon}{8\pi}{\bi E}^2\bigg ] ,
\en 
where  $C$ is   a constant  and 
\bea 
g(n) &=& f(n)- \mu_\infty n + p_\infty\nonumber\\
&=& f(n)-f_\infty- \mu_\infty (n-n_\infty), 
\ena 
with $f_\infty$ being  the Helmholtz 
free energy density 
  far from the ion.  The $\Delta\Omega$ 
will be called the solvation free energy.   
Here  $g$ and 
$\p {g}/\p n$ tend to zero 
at $n \rightarrow n_\infty$ 
so that we have  
$g \cong (n-n_\infty)^2/2n_\infty^2 K_T(n_\infty)$  
far from the ion.
It is worth noting that if 
 the electrostriction is neglected 
or $n= n_\infty$ in the whole space,  
$\Delta\Omega$ consists of the  
electrostatic free energy $F_e$ only,  leading to  the   
Born contribution \cite{comment}, 
\be
\Delta\Omega_{{\rm Born}} = \int d{\bi r} \frac{\ve_{\infty}}{8\pi} 
\bigg(\frac{Ze}{\ve_{\infty}r^2}\bigg)^2=
\frac{Z^2e^2}{2\ve_\infty R_{\rm B}},   
\en 
where $\ve_\infty=\ve(n_\infty)$.    
Here $R_{\rm B}=R_{\rm i}$ if 
the space integral is outside the ion  
($r>R_{\rm i}$) \cite{cutoff} with the upper bound 
 pushed to infinity.  
However, this  contribution  
much overestimates $F_e$ 
when $n(r)$ approaches a  
liquid density close to the ion 
and tends to a smaller (gas or supercritical) density 
far from it. 
In such cases  $F_e$  is  of 
order $Z^2e^2/2\ve_\ell R_{\rm B}$ with 
$\ve_\ell$ being the dielectric constant in liquid \cite{OK}, 
which is much smaller than $\Delta\Omega_{{\rm Born}}$ in 
Eq.(2.17) for  $\ve_\ell \gg \ve_\infty$.

In Eq.(2.15) $F_e$ arises from 
the ion-dipole  
interaction among the ion and 
the solvent molecules. However, 
there is also a  repulsive  interaction 
among them.  To account for it   
we may include  a  potential $v_{\rm is}(r)$ to obtain 
another form of the grand potential increase given by        
\be 
\Delta \Omega'  = \Delta \Omega + \int d{\bi r}
 v_{\rm is}(r) n(r)  ,
\en 
as in the density functional theory in Ref.6.  
The potential   $v_{\rm is}(r)$ should grow strongly 
close to the ion center. 
In this paper we use the truncated Lenard-Jones 
potential $v_{\rm is}$ defined by 
\be  
v_{\rm is}(r) = 
\epsilon_{\rm vw}
[(\sigma_{\rm is}/r)^{12} - (\sigma_{\rm is}/r)^6] 
\quad (r<\sigma_{\rm is}),  
\en 
and $v_{\rm is}(r)=0$ for $r>\sigma_{\rm is}$  
with   $\sigma_{\rm is}= \sigma/2 + R_{\rm i}$. 
We  will show numerical  results 
starting with either  Eq.(2.15) or  Eq.(2.18).

In this paper the coefficient $C$ 
of the gradient term in 
Eq.(2.15) will be set equal to    
\be 
C=10\sigma^5 k_{\rm B}T. 
\en  
As will be shown in Appendix A, 
this  choice of $C$ is 
consistent with a calculation   of the surface tension 
$\gamma$ for water in Ref.23. The thermal correlation length 
$\xi= (Cn^2K_T)^{1/2}$ is then about $\sigma$ 
for $T/T_c=0.675$ and   about  $2.2\sigma$ 
for $T/T_c=0.928$ in liquid on the coexistence curve.

\subsection{Solvation in gas }

When an ion solvates the surrounding 
fluid in stable or  metastable one-phase states, 
we may  determine  the density  profile 
$n=n(r)$ by requiring    
the extremum condition
${\delta \Delta \Omega}/{\delta n}=0$ 
or ${\delta \Delta \Omega'}/{\delta n}=0$ 
expressed  as 
\be 
{\tilde \mu}   -C\nabla^2n -\mu_\infty=
{\epsilon_{\rm vw} A\sigma}\frac{\ve'\theta^2}{\ve^2 r^4},   
\en  
where $\tilde \mu$ is  equal to 
$\mu$ in  Eq.(2.10) for ${\delta \Delta \Omega}/{\delta n}=0$ 
or to $\mu+ v_{\rm is}$ for ${\delta \Delta \Omega'}/{\delta n}=0$. 
The right hand side is $\mu_e$ in Eq.(2.6).
We  introduce  the dimensionless 
strength of the electric field, 
\be 
A= \frac{Z^2e^2}{8\pi
\sigma\ep_{\rm vw}}= \frac{4}{27\pi}
\frac{Z^2e^2}{\sigma k_{\rm B}T_c},
\en 
using $\epsilon_{\rm vw}$ in Eq.(2.8)  or $T_c$ in Eq.(2.13). 
Here we estimate $A=3.9Z^2$ 
for water. 
As shown in our previous work\cite{OK}, the gradient term becomes 
negligible in Eq.(2.21) far from the ion. 
The long distance behavior of 
the deviation $\delta n= n-n_\infty$ 
 far from the critical point   is 
\be 
\delta n \cong D_\infty {\epsilon_{\rm vw} A\sigma}/{ r^4}\quad 
(r \gg \sigma)
\en 
with $D_\infty$ being the value of 
$n^2K_T {\ve'}/{\ve^2}$ far from the ion. 
The solvation profile is more complicated 
near the critical point due to  the growing 
correlation length\cite{OK}.

In  numerically solving  Eq.(2.21) 
the ion radius will be assumed to be given by  
\be 
R_{\rm i}=0.3\sigma. 
\en  
which is taken to be considerably smaller 
than $\sigma$ and then $\sigma_{\rm is}= 0.8\sigma$ in Eq.(2.19).  
In the literature\cite{Is,RMarcus} 
the ion radius  has  been estimated for various ions in water. 
For example, in Table V in 
Ref.5,  the  {\it bare} ion radius 
 is approximately 
 $0.68$${\rm \AA}$ for Li$^+$, 
$0.95$${\rm \AA}$ for Na$^+$, 
$1.69$${\rm \AA}$ for Cs$^{2+}$, and  
$0.5$${\rm \AA}$ for Al$^{3+}$, which  
 are considerably smaller than the corresponding 
hydration shell radius. 
The bare radius in 
Ref.5 corresponds to $R_{\rm i}$ in our theory.

In Fig.3 we show the  normalized density profile   
$
\phi(r)=v_0 n(r)
$
 around an ion 
in the gas phase case $\phi_\infty= 
v_0n_\infty=0.049$  for $A=5$ and 50 by minimizing  $\Delta\Omega'$ 
in Eq.(2.18) including $v_{\rm is}$. Note that 
$n(r)$ tends to zero rapidly at small $r$.  
Similar density profiles 
were calculated in the previous 
studies \cite{Kusaka,Zeng,Elena}. On the other hand, if 
we minimize $\Delta\Omega$ in  Eq.(2.15) without $v_{\rm is}$,
    $n(r)$  tends to a positive constant $n(0)$ 
at small $r < R_{\rm i}$, as for the concentration 
in our previous work \cite{OK}. 
  We show that 
the solvation density profiles change continuously 
close to the coexistence curve 
with a change in the temperature difference,  
\be 
\delta T=T -T_{\rm cx}, 
\en 
where $T_{\rm cx}$ is the coexistence temperature 
at given density.  This means that  there is no 
 discontinuous change in the solvation profile and 
the free energy near the coexistence curve \cite{Fukuda}.

Let us consider the condition of strong  
electrostriction in the gas phase, where $n(r)$ 
assumes a liquid density close to an ion. 
Note that the chemical potential 
on the left  hand side of Eq.(2.21) is of order 
$k_{\rm B}T/(1-v_0n)$ and $\ve'$ is of order 
$ v_0 \ve$ in liquid states 
with $k_{\rm B}T \sim \epsilon_{\rm vw}$ and 
$n \sim v_0^{-1}$. Thus  
strong solvation condition is given by  
\be 
A (\sigma/R_{\rm i})^4/\ve_{\rm \ell} \gg 1,
\en 
where $\ve_{\rm \ell}$ is the dielectric 
constant in liquid.  The degree of solvation 
strongly increases with decreasing 
the relative ion size $R_{\rm i}/\sigma$, 
as is well-known in the literature \cite{Is}.  
The shell  radius $R_{\rm shell} $  may  be 
defined such that  
$ n (r)$ is on the order of the liquid density 
within the shell region $r < R_{\rm shell} $. 
Then we find  
$A (\sigma/R_{\rm shell})^4/\ve_{\rm \ell} \sim 1$ to 
obtain   
\be 
R_{\rm shell} 
\sim \sigma (A/\ve_{\rm \ell})^{1/4},
\en 
which can exceed  $\sigma$ for 
$A/\ve_{\rm \ell} \gs 1$.

\subsection{Solvation on the coexistence curve}

Due to the large  density variation around an  ion 
in gas,   the  effect of electrostriction is   
more marked   in gas than in liquid. 
As a result, the solvation free energy    
is larger in gas than in liquid on the 
coexistence curve.   We show 
$\beta\Delta \Omega$ without $v_{\rm is}$ 
in Fig.4 and $\beta\Delta \Omega'$ with $v_{\rm is}$ 
in Fig.5 for the two phases along the coexistence curve. 
We notice that  
$\beta\Delta\Omega'$ 
 even exceeds 1000 on the right panel of Fig.5. 
 This is because $\ve (n)$ decreases  
from a liquid value to a gas value in the region 
$r<\sigma_{\rm is}$ in the presence of  $v_{\rm is}$. 
As a result,  the integration of 
the electrostatic energy density in the region 
$R_{\rm i}< r< \sigma_{\rm is}$ becomes of 
order $\Delta\Omega_{\rm Born}$ in Eq.(2.17) 
($\sim 10^3 k_{\rm B}T$ 
for $A=50$).

In the next section we 
 shall see that  relevant in nucleation is 
the difference between the gas value and 
the liquid value of the solvation free energy,  
\bea 
\Delta_{\rm sol} &=& 
(\Delta\Omega)_{n=n_{\rm g}} -(\Delta\Omega)_{n=n_{\rm \ell}} 
\quad {\rm without}~ v_{\rm is} 
\nonumber\\
&=& 
(\Delta\Omega')_{n=n_{\rm g}} -(\Delta\Omega')_{n=n_{\rm \ell}} 
\quad {\rm with}~ v_{\rm is}  .  
\ena 
Here the grand potential increases,  $\Delta\Omega$ 
and $\Delta\Omega'$,
are calculated using the inhomogeneous solutions 
of Eq.(2.21),  $n=n_{\rm g}(r)$ for  gas 
or  $n=n_{\rm \ell}(r)$ for liquid on the coexistence curve. 
For the case with $ v_{\rm is}$ we define  the grand-potential density,  
\be 
\omega'(r)= g+ v_{\rm is}+ \frac{C}{2}|\nabla n|^2+ 
\frac{\ve}{8\pi} {\bi E}^2.
\en 
Then  we have 
\be 
\Delta_{\rm sol} 
= 4\pi\int_0^\infty dr  r^2[\omega'_{\rm g}(r)-
\omega'_{\rm \ell}(r)], 
\en  
where $n=n_{\rm g}(r)$ in  
$\omega'_{\rm g}$ and $n=n_{\rm \ell}(r)$ in  
$\omega'_{\rm \ell}$.  The two terms, 
$\omega'_{\rm g}(r)$ and $
\omega'_{\rm \ell}(r)$, in the integrand mostly cancel 
 in the region $r<\sigma_{\rm is}$. 
In Fig.6 we show $\beta\Delta_{\rm sol}$ 
on the coexistence curve for $A=5$ (left) 
and $50$ (right).  As explained below Eq.(2.17),   
 the Born approximation  
much overestimates 
$\Delta_{\rm sol}$ in strong solvation.

\section{Ion-induced nucleation  in gas}
\setcounter{equation}{0}

\subsection{Critical droplet and KO approximation}

In nucleation theory \cite{Onukibook,OxtobyReview,De96} 
crucial is the free energy increase 
 to create a critical droplet, which will
 be written as $W_c$ and 
 will be called the nucleation barrier.  
Its definition for ion-induced nucleation 
will follow in Eq.(3.3) below. 
The nucleation rate $J$ is 
the birth rate  of droplets of the new phase 
with radius  larger than 
the critical radius $R_c$ 
emerging in a metastable state per unit volume. 
It is approximately  
given by 
\be 
J= J_0 \exp(- \beta W_c),
\en   
independently of the details of 
the dynamics.   Here 
 $J_0$ is a microscopic coefficient 
(of order $n/\tau_m$ with $\tau_m$ being 
a microscopic time away from the critical point)  
 and $\beta=1/k_{\rm B}T$. A unique aspect  of 
nucleation in  gaseous  polar fluids  is that   
the  solvated region around an ion  
 can serve as a  seed of 
a nucleating liquid droplet in a metastable   state. 
This leads to   a  considerable 
decrease in $W_c$ and a dramatic increase in $J$ 
in the presence of a small amount of ions.

Mathematically, 
the extremum condition 
(2.21) holds also for 
the critical droplet profile, 
\be 
n=n_{\rm cri}(r),
\en 
which assumes liquid values 
for $r\ls R_c$ and 
tends to a metastable 
gas density $n_\infty$  at large $r \gg R_c$. 
This solution  of Eq.(2.21) is  unstable; that is, 
if we superimpose a small  density 
perturbation inside  a critical droplet,  
$\Delta\Omega$ (or $\Delta\Omega'$) decreases\cite{grow}. 
In Fig.7   with $v_{\rm is}$ 
being included,  
the normalized density $\phi(r)=v_0n_{\rm cri}(r)$ 
for a critical droplet  
is displayed around an ion  with $A=50$  for  $T/T_c=0.928$ (left) 
and  $0.675$ (right). 
 The dotted and broken lines  represent  
 the solvation profile  in  a 
 metastable gas  
and  that in a  stable liquid, respectively.   
Here $\delta T/T_c$ is $-0.011$ (left) 
and $-0.046$ (right), so $R_c$ is larger 
for the former case.

The  nucleation barrier 
$W_c$ is then written as 
\bea 
W_c &=& 
(\Delta\Omega)_{n=n_{\rm cri}} -
(\Delta\Omega)_{n=n_{\rm g}} 
\quad {\rm without}~ v_{\rm is} 
\nonumber\\
&=& 
(\Delta\Omega')_{n=n_{\rm cri}} 
-(\Delta\Omega')_{n=n_{\rm g}} 
\quad {\rm with}~ v_{\rm is}  ,  
\ena 
where   $n_{\rm cri}(r)$ 
is the critical-droplet solution 
and $n_{\rm g}(r)$ is the inhomogeneous solution 
 for the metastable gas.  
In terms of $\omega'$ in Eq.(2.29) 
we have 
\be 
W_c= 4\pi \int_0^\infty dr r^2[\omega'_{\rm cri}(r)-
\omega'_{\rm g}(r)], 
\en  
for the case with 
$ v_{\rm is}$.  Here 
 $n=n_{\rm cri}(r)$ in  
$\omega'_{\rm cri}$ and $n=n_{\rm g}(r)$ in  
$\omega'_{\rm g}$.  
The  numerical calculation of  $W_c$ 
in this manner will be  
referred to as KO approximation. 
In Fig.7 the calculated  
$\beta W_c$  is  $50.58$ (left) and  $101.24$ (right).
In Fig.8 we show the dimensionless quantity  
$4\pi \beta \sigma r^2[\omega'_{\rm cri}(r)-
\omega'_{\rm g}(r)]$ for the case of 
$T/T_c=0.928$ (left) in Fig.7, which is 
the integrand in Eq.(3.4) multiplied by $\beta\sigma$. 
It  consists of a negative part 
within the droplet $r \ls R_c$ and a positive surface part 
at $r\sim R_c$, and its integral with respect to $r/\sigma$ 
is equal to $\beta W_c\sim 50$.

\subsection{KO' approximation} 

In the KO approximation for $W_c$ we need to calculate 
$n_{\rm cri}(r)$ for each given set of the 
parameters $\delta T$, $A$, and $R_{\rm i}$.  
 We here propose a simpler 
(KO') approximation valid for  shallow quenching. 
That is, we set   
\be 
W_c=  -\Delta_{\rm sol}+W_c^0. 
\en 
where $\Delta_{\rm sol}$ 
is the constant independent of $R$ 
 defined  by Eq.(2.28) 
and $W_c^0$  is the nucleation free energy 
in the absence of  ions 
or for homogeneous nucleation.  
The first term 
 arises from the difference between  
the solvation free energy  
in gas and   liquid, which 
may be equated to the values  on the coexistence curve 
for shallow quenching  
as indicated by Fig.3.   
  In  curves in Figs.9 and 10 to follow, 
 $W_c^0$ will be  numerically obtained from  
 Eq.(2.21) for pure fluids 
or for $Ze=0$ and $v_{\rm is}=0$.

For large droplets ($R\gg \xi$) 
we may propose the following 
 droplet free energy,   
\be 
W_{\rm KO}(R) 
=-\Delta_{\rm sol}+ 4\pi \gamma R^2 -\frac{4\pi}{3}\delta\mu R^3 .
\en 
The last two terms    have the meaning of  the 
minimum work  needed to form 
a droplet with radius $R$ in homogeneous nucleation \cite{De96}. 
Namely, the second term is the surface part with
 $\gamma$ being the surface tension. 
The third term 
with $\delta\mu=-g(n_\ell)$ 
is the bulk part, which is negative or 
 $\delta\mu>0$  in metastable states.  Notice 
that the dotted curve in Fig.8 represents 
$-4\pi (\beta\sigma\delta\mu) r^2$. 
See Appendix B for 
more discussions on $\delta\mu$. 
This $W_{\rm KO}(R)$ is maximized 
at the Kelvin radius, 
\be 
R_{\rm K}= 2\gamma/\delta\mu,  
\en 
which is the critical radius in homogeneous nucleation.  
The maximum value is the nucleation barrier, 
\be 
W_c\cong -\Delta_{\rm sol}+ 
\frac{4}{3} \pi \gamma R_{\rm K}^2 .  
\en 
In this simple case 
Eq.(3.1) may be rewritten as 
\be 
\delta\mu= (16\pi\beta\gamma^3/3)^{1/2} 
/[ \ln (J_0/J)+ \beta \Delta_{\rm sol}]^{1/2}.  
\en 
Appreciable droplet formation 
is observed for sufficiently large nucleation rate $J$ 
 dependent  on the experimental method 
($\sim 1$$/$cm$^3$sec, say) 
and the above formula 
determines the so-called cloud point 
of nucleation \cite{Onukibook}.

\subsection{Thomson approximation}

Since Thomson's work  \cite{Thomson},  
the droplet free energy for large $R$ 
has been assumed to be of the form 
\cite{Seto,Nad,NewPRL}, 
\be 
W_{\rm T}(R)= -\frac{\alpha}{R_{\rm i}}+\frac{\alpha}{R}   
+4\pi \gamma R^2 -\frac{4\pi}{3}\delta\mu R^3  , 
\en 
where  $\alpha$ is written 
in terms of the dielectric constant 
in gas ${\ve_{\rm g}}$ 
 and  that in  liquid ${\ve_{\rm \ell}}$ as 
\be 
\alpha= \frac{1}{2} {Z^2e^2} (
{1}/{\ve_{\rm g}}- {1}/{\ve_{\rm \ell}} ).    
\en 
The  first two terms in Eq.(3.10) 
represent  the difference of the electrostatic 
free energy $F_e$ between the two phases,   
\be 
(\Delta F_{e})_{\rm Born}=
- \int_{R_{\rm i}}^{R} dr \frac{\alpha}{r^{2}},  
\en 
where the upper bound of the integration 
is $R$ since 
 $\ve(n)$  changes from 
$\ve_{\rm g}$ to ${\ve_{\ell}}$ 
at $r = R$  with increasing $r$.  
The first term in Eq.(3.10) arises from the lower bound 
and turns out to be 
the Born approximation of 
$-\Delta_{\rm sol}$ in Eq.(2.28).

Thomson \cite{Thomson} found  that the sum of the 
second and third terms on the right hand 
side of Eq.(3.10) is  minimized at 
the  Rayleigh radius \cite{Ra},   
\be 
R_{\rm R}= (\alpha/8\pi \gamma)^{1/3} .
\en 
He concluded that an ion in gas 
is surrounded by a spherical 
liquid  region with radius $R_{\rm R}$.  
However, as estimated by Thomson himself, 
$R_{\rm R}$ is only about $3$$\rm \AA$ 
for an ion with $Z=1$ in room-temperature 
water vapor, while  Eq.(3.10) should be  valid only 
 for large $R$.  For not very large $Z$ we can see that  
$R_{\rm R}$ cannot  exceed even the 
interface thickness    at
 any temperatures \cite{critical}. 
In addition, use of the surface 
tension makes his argument applicable only 
in the region  $\delta T\le 0$.    
As shown in Fig.3,  the solvation density profiles of small ions 
should be  continuous through  the coexistence curve. 
Obviously,   the 
 second term ($\propto R^{-1}$) in Eq.(3.10) 
 is  negligible for 
$R \gg R_{\rm R}$  and the Thomson theory is not  well justified 
(if applied to  microscopic ions).

We explain how the nucleation free energy $W_c$ is  calculated  
in the Thomson theory \cite{Seto}. 
The extremum condition 
$\p W_{\rm T}(R)/\p R=0$ 
becomes $x^3(1-x)= \alpha^*$ with $x= R/R_{\rm K}$, where 
\be 
\alpha^*= {\alpha}/{ 8\pi \gamma R_{\rm K}^3}= 
({R_{\rm R}}/{R_{\rm K} } )^3 . 
\en  
We can see that $W_{\rm T}(R)$  has a maximum  
$W_{\rm max}$ and a minimum  
$W_{\rm min}$ for 
\be  
\alpha^*<  {27}/{256}\quad   {\rm or} \quad 
{R_{\rm R}}/{R_{\rm K} } < {3}/{4^{4/3}}.
\en  
Since $R_{\rm R}$ is microscopic for not large $Z$,  
we have $\alpha^* \ll 1$ for shallow quenching. Obviously, 
the radius giving the maximum 
is nothing but  the critical radius $R_c$. 
For small $\alpha^*$, $R_c$ behaves as 
$
R_c= R_{\rm K}(1- \alpha^*+\cdots),   
$    
while the radius giving the minimum is nearly equal to 
$R_{\rm R}$.  
If  Thomson's idea is extended, 
a preexisting {\it liquid droplet}  
with radius $R_{\rm R}$ grows into a larger liquid 
 droplet in nucleation. 
Within his theory, the nucleation barrier is  
 given by \cite{Seto}  
\bea 
W_c &=& W_{\rm max}-W_{\rm min}\nonumber\\ 
&=& \frac{4}{3}\pi \gamma R_{\rm K}^2(1+   8\alpha^*)-
12\pi\gamma  R_{\rm R}^{2}+\cdots .
\ena 
In the first line  the first 
 term  in Eq.(3.10)  is canceled  to vanish. 
The second line is an  expansion in powers of 
$(\alpha^*)^{1/3}$  for small  
$\alpha^*$. This approximation   was adopted in 
Ref.3, but  it 
 is not well justified for small 
$R_{\rm R}$ and hence for small ions.

\subsection{Numerical results for $W_c$}

Without  $v_{\rm is}$ in Fig.9 
and with  $v_{\rm is}$ in Fig.10, 
the normalized nucleation barrier  
$\beta W_c$  is plotted as a function of  $A$ 
at $T/T_c=0.928$ (left)  and 
$0.675$ (right).   The profiles of the critical droplets 
are shown in Fig,7 at $A=50$ for these temperatures. 
The points  ($\bullet$) (KO) are obtained from 
the profile of the 
critical droplet satisfying   Eq.(2.21). 
The solid line represents  
$\beta W_c$ in Eq.(3.5) (KO'), 
while the broken line represents 
$\beta W_c$ in Eq.(3.16) (Thomson). 
It is remarkable that  there is  
no essential difference  between Fig.9 
without  $v_{\rm is}$ and 
Fig.10 with  $v_{\rm is}$.  
(i) For $T/T_c=0.928$,   
 good agreement is obtained 
between the KO  result  
and the KO'  result 
because of relatively large $R_c$, while the Thomson result 
is larger by about 10. 
For  example,  for $T/T_c=0.928$ and  $A=50$ in Fig.10, 
$\beta W_c$ 
is given by  $50.58$ (KO), 
$50.71$ (KO'), and $61.59$ (Thomson). 
(ii) For $T/T_c=0.675$,    
the KO values of $\beta W_c$ 
agree with the KO' values  for $A \ls 10$,  
but become considerably 
larger for  $A \gs 20$.  In fact,
  $\beta W_c$  is given by 
$101.24$ (KO), $77.80$ (KO'), 
and $80.96$ (Thomson) for   
 $T/T_c=0.675$.  
In this   lower temperature case   
$R_c$ is of order $R_{\rm shell}$, 
as shown in Fig.4,  and hence 
the expression of $W_{\rm T}(R)$ in Eq.(3.10) should  
not be a good approximation.

For these two temperatures $\beta W_c$ decreases 
with increasing $A$.   At small $A$($\ls 5)$ 
the curves are rather steep, mainly due to 
the small size of $R_{\rm i}/\sigma=0.3 $ in   
Eq.(2.24).    In the density functional 
theory Kusaka {\it et al.} \cite{Kusaka} 
 calculated $W_c$ for $R_{\rm i}=\sigma$ 
as a function of 
 another   parameter 
$\chi$ ($=Ze p_0/\sigma^2k_{\rm B}T_c$ 
with $p_0$ being the dipole moment 
of a polar molecule)  and found that it is 
rather weakly dependent on $\chi$ 
for small $\chi$.

\section{Summary and concluding remarks} 
\setcounter{equation}{0}

In summary, we have presented a  simple  theory 
of ion-induced nucleation in polar one-component   
fluids, where the effect of 
electrostriction is crucial. We summarize our main 
results.\\
(i) We use  a simple continuum model, 
where the electrostatic free energy 
is given by Eq.(2.5) 
and the grand potential increase due to an ion 
is given by  Eq.(2.15) or  Eq.(2.18). 
Particularly, 
 we use the Harris-Alder formula (2.7) 
for the dielectric constant as a function of the density 
and the van der Waals form 
for the Helmholtz free energy density in Eq.(2.8). 
The density profiles around an ion 
are numerically obtained as illustrated in Figs.3 
and 7.\\
(ii)  We have  proposed the droplet  
free energy  $W_{\rm KO}(R)$ in Eq.(3.6) 
for a large liquid droplet in a metastable gas 
with an ion at its center. It is remarkable that 
the difference of solvation in   
gas and in liquid gives rise to the 
 negative background 
contribution $-\Delta_{\rm sol}$, 
which much favors nucleation around an ion 
 in a metastable gas.\\
(iii)  
The  nucleation barrier 
$W_c$ has been calculated in three manners. 
In our Ginzburg-Landau  scheme we have directly 
 sought  an unstable solution of 
Eq.(2.21) to obtain 
the points ($\bullet$) (KO) in Figs.9 and 10. 
It may also be  calculated more 
approximately using  Eq.(3.5) (KO')
 or  the first line of Eq.(3.16) (Thomson). 
They yield the solid lines and 
 the dotted lines in  Figs.9 and 10, 
respectively. We have found that the KO' result 
is close to the directly 
calculated result (KO) for relatively 
large critical radius 
$R_c$ if the background is  chosen to be 
$-\Delta_{\rm sol}$.\\
(iv) We have presented 
the density profiles in Figs.4 and 7, 
which go to zero at
 the  ion center  owing to   
the pair potential $v_{\rm is}$. 
However, we have obtained 
 essentially the 
same results for the free energy barrier $W_c$ 
in Fig.9 without  $v_{\rm is}$ 
and in  Fig.10 with  $v_{\rm is}$.

Finally we make some remarks.\\
(i) We have used the Ginzburg-Landau theory 
with the simple gradient 
free energy. However, far from the critical point 
as in the case of $T/T_c=0.675$,  
the density functional theory 
\cite{Kusaka,OxtobyReview,Oxtoby} 
with a nonlocal pair interaction 
would give more improved  results.  
\\
(ii) Because of its simplicity 
we have  used the macroscopic linear dielectric 
formula (2.7) even very close to the ion.  
This has  been the approach 
 in the previous continuum theories\cite{Thomson,Born}, but 
it is not well justified.  In particular, 
 we should examine 
 the effect of nonlinear  dielectric saturation 
in solvation, which gives rise to   
 a decrease in  the effective dielectric constant close to the ion 
\cite{Abraham,nonlinear}.\\ 
(iii) The difference  $\Delta_{\rm sol}$   of the 
 solvation  free energies in Eq.(2.28) 
is a relevant parameter  generally in ionic solutions 
in two-phase states, although the present 
 work has treated the case of 
a single ion. For example,  it gives rise to 
ion-density differences inside and outside 
 a  wetting  layer.\\ 
(iv) Obviously  $W_c$   
tends to zero on approaching the critical point. 
However, this is the result for a single ion. 
We point out that  even a  small concentration of  ions
can drastically  alter 
  the  phase behavior of  near-critical 
polar fluids \cite{OK}.\\ 
(v) We can also construct 
a theory of ion-induced nucleation in 
binary fluid mixtures \cite{OK}. \\

\vspace{2mm} 
{\bf Acknowledgments}
\vspace{2mm}

 We would like to thank 
M. Anisimov for valuable discussions and 
providing us a copy of 
the article by Fernandez {\it et al}.\cite{Fernandez} 
This work was  supported by 
Grants in Aid for Scientific 
Research 
and for the 21st Century COE "Center 
for Diversity and Universality in Physics" from the Ministry of Education, 
Culture, Sports, Science and Technology  of Japan.

\vspace{10mm} 
{\bf Appendix A}\\
\setcounter{equation}{0}
\renewcommand{\theequation}{A.\arabic{equation}}

Here we consider  the surface tension 
 $\gamma$ in the absence of ions 
in our model in the mean field theory 
\cite{Onukibook}.  To have a planar interface 
we assume that the fluid is 
on the coexistence curve 
or the pressure is given by 
  $p=p_{\rm cx}(T)$ with the temperature below $T_c$. 
The grand-potential density 
\be 
g(n)= f(n)- \mu_{\rm cx}n+ p_{\rm cx}
\en 
is minimized at $n=n_{\rm g}$ and $n_\ell$ 
on the coexistence curve with the minimum value being zero. 
For    our van der Waals  free energy $f$ in Eq.(2.8) 
and  the gradient free energy 
in Eq.(2.15),  a planar interface profile 
$n=n(x)$ changing along 
the $x$ axis satisfies $\p g/\p n= C\p^2 n/\p x^2$, 
yielding  $g= C(\p n/\p x)^2/2$. 
Then $\gamma$ is expressed as 
\bea 
\gamma &=&  \int_{-\infty}^\infty  dx 
[g+  C(\p n/\p x)^2/2]\nonumber\\
&=& (2C)^{1/2} \int_{n_{\rm g}}^{n_\ell} dn 
[ g(n)]^{1/2} .
\ena 
In the vicinity of the critical point it follows 
the power law behavior,  
\be 
\gamma=  C_{\rm s} 
\epsilon_{\rm vw}(1-T/T_c)^{3/2}/\sigma^2,   
\en 
where $  C_{\rm s} $ is a constant. 
Widely  in the range  
$T/T_c \gs 0.6$,   however, we have found  that 
 $\gamma$ numerically  calculated from the second line 
 of Eq.(A.2) with Eq.(2.13) can be fitted to 
Eq.(A.3)   within a few percents if we set   
\be 
C_{\rm s}=3.0.   
\en  
With this   $C_s$ 
 Eq.(A.2) yields   
$\gamma \cong  235 (1-T/T_c)^{3/2}$ dyn$/$cm 
where  we use the values of 
 $\epsilon_{\rm vw}$ 
and $\sigma$  given below Eq.(2.13) for water.

 Experimentally\cite{Kiselev}, 
 the surface tension of water behaves 
as $215(1-T/T_c)^{2\nu}$ dyn$/$cm 
with $\nu \cong 0.63$ in the range $1-T/T_c \ls 0.1$.
However, the calculated $\gamma$ 
 fairly agrees  with the experimental  surface tension of 
water far  from  the critical point.  For example, at 
$T/T_c=0.675$, our $\gamma$ 
is      42.5 dyn$/$cm, while  
the experimental value is 
44.6 dyn$/$cm.  Since $C_{\rm s} \propto C^{1/2}$, 
our choice of $C$ in Eq.(2.21)  is  justified.\\

\vspace{2mm} 
{\bf Appendix B}\\
\setcounter{equation}{0}
\renewcommand{\theequation}{B.\arabic{equation}}

We examine the behavior of 
$\delta\mu=-g(n_{\ell}) $ in  Eq.(3.6) first   fixing  $T$ 
below $T_c$. 
From  Eq.(2.16)
it  becomes      
\bea
\delta\mu &=&p_\ell-p_\infty + 
n_{\rm \ell}( \mu_\infty- \mu_{\ell}) \nonumber\\
&=&
- (f(n_\ell)-\mu_{\rm cx}n_\ell+p_{\rm cx}) \nonumber\\
&&+n_{\rm \ell}( \mu_\infty- \mu_{\rm cx})
 - (p_\infty -p_{\rm cx})
 .
\ena 
The quantities with the subscript $\ell$, 
$\infty$, and $\rm cx$ 
 are those 
within  the liquid droplet, 
 in the metastable gas, and on the coexistence curve, respectively.
See Ref.28 for  a   
derivation of the first line of Eq.(B.1) from 
statistical-mechanical  principles.  
In the second line,  the first term 
is negligible since $n_\ell$ is close 
to the liquid density on the coexistence curve. 
In the limit  $p_{\rm cx}-p_\infty \rightarrow 0$,
 the Gibbs-Duhem 
relation gives  
\be 
\delta\mu \cong 
(\Delta n/n_g) (p_\infty- p_{\rm cx}),
\en  
where $\Delta n=n_\ell-n_g$. 
This relation  is useful near the critical point. 
For dilute gas we have 
$p\cong k_BT n$ and 
$\mu_\infty- \mu_{\rm cx}  
= \int_{p_{\rm cx}}^{p_\infty} dp n(p)^{-1} 
\cong  k_{\rm B}T \ln (p_\infty/p_{\rm cx})$ at fixed $T$.   
If $n_{\rm g}\ll n_{\rm \ell}$, 
the second term  is much smaller than the first 
on the right hand side of Eq.(3.10),  
leading to the well-known 
expression \cite{Seto,Mora},   
\be 
\delta\mu\cong 
 k_{\rm B}T \ln(p_\infty/p_{\rm cx}).
\en

In real experiments the temperature 
$T_\infty$ in gas  (written as $T$ in the above discussion) 
may also  be changed.  Let us consider a 
reference state at temperature $T_0$ and pressure 
$p_0= p_{\rm cx}(T_0)$ on the coexistence curve. 
 In  such cases  
we should  replace 
$p_\infty-p_{\rm cx}(T_0)$  in the above formulae by 
$p_\infty-p_{\rm cx}(T_0) - 
(\p p/\p T)_{\rm cx}(T_\infty-T_0)\cong 
p_\infty-p_{\rm cx}(T_\infty)$ 
for shallow quenching,   
where $(\p p/\p T)_{\rm cx}$ is the derivative 
along the coexistence curve 
\cite{Onukibook}.  In Fig.2,  for example, 
the  temperature is changed 
across the coexistence curve at fixed 
average  density $n_0$ or at fixed system volume,  
 where we may set $T_0=T_{\rm cx}(n_0)$. 
This means 
$p_\infty-p_{\rm cx}(T_0)=(\p p/\p T)_n(T_\infty-T_0)$ 
in the gas, leading to  
$\delta\mu= [\Delta n (\p n/\p T)_{\rm cx}
/n_g^2K_T](T_0-T_\infty)$ in terms of the compressibility 
$K_T$ in the gas. Here 
 use is made of the thermodynamic relation 
 $(\p p/\p T)_n- (\p p/\p T)_{\rm cx}= 
 -(\p n/\p T)_{\rm cx}/nK_T$ on the coexistence curve
(see Eq.(2.2.39) of Ref.$[21]$).


\end{multicols}


\newpage 
\noindent
\centerline{FIGURE CAPTIONS}\\

\noindent
FIG. 1.  Static dielectric constant $\ve$ as a function of $\phi=v_0n$  
at $T/T_c=0.675$ and 0.928 
on the basis of the Harris-Alder formula 
(2.7) using data of water in Ref.17. 
The dotted part corresponds   to the two-phase region 
in the van der Waals model 
and is given by $0.0358<\phi<0.729$ for $T/T_c=0.675$  
and by $0.168<\phi<0.519$ for $T/T_c=0.928$. 
\vskip 4mm

\noindent
FIG. 2.  Phase diagram in the plane of 
$T/T_c$  and $\phi=v_0n$  with 
coexistence curve (solid line) and  spinodal 
curve (dotted line). Use is made of the van der Waals model. 
 At the points ($\bullet$) 
numerical calculations  are performed in this work. 
\vskip 4mm

\noindent
FIG. 3.  Normalized density $\phi(r)=v_0 n(r)$ around 
an ion in the gas phase 
($\phi_\infty=v_0n_\infty= 0.049$)  
close to the coexistence curve 
($T_{\rm cx}= 0.721$) 
for $A=5$ and 50.  The data are for the three 
points with $\phi=0.049$ in Fig.2. 
The curves  for $T/T_c=0.759$ represent  stable profiles, 
while the others  metastable ones. However, 
for  $T/T_c=0.591$ and $A=50$,  
there is no metastable solution 
with  $\lim_{r\rightarrow \infty} n(r)=n_\infty$. 
\vskip 4mm

\noindent
FIG. 4.  Normalized  solvation free energy 
$\beta \Delta\Omega$ without $v_{\rm is}$ 
 around an ion 
 for gas and liquid  on the coexistence curve 
for $A=5$ (left)  and 50 (right). 
\vskip 4mm

\noindent
FIG. 5.  Normalized  solvation free energy 
$\beta \Delta\Omega'$ with $v_{\rm is}$ 
 around an ion 
 for gas and liquid  on the coexistence curve 
for $A=5$ (left)  and 50 (right). 
\vskip 4mm

\noindent
FIG. 6.  Normalized solvation free energy  
difference $\beta\Delta_{\rm sol}$  defined by Eq.(3.5) 
along the coexistence curve 
with and without  
$v_{\rm is}$ for $A=5$ (left)  and 50 (right).
\vskip 4mm

\noindent
FIG. 7.   Normalized density $\phi(r)$ 
around an ion with $A=50$ in a metastable 
gas  with $\phi_\infty=0.18$ (dotted line) 
and in the corresponding  liquid 
 (broken line)  at $T/T_c=0.928$ (left) 
and  $0.675$ (right).  Also shown 
 is the critical droplet obtained from Eq.(2.19). 
Here $v_{\rm is}$ is present and  $n(r)$ tends to zero 
at small $r$. 
\vskip 4mm

\noindent
FIG. 8.    
$4\pi \beta \sigma r^2[\omega'_{\rm cri}(r)-
\omega'_{\rm g}(r)]$, whose 
integration with respect to  $r/\sigma$ gives 
$\beta W_c$ as in Eq.(3.4).  The dotted line 
represents $-4\pi(\beta \sigma \delta\mu )r^2$ 
where $\delta\mu$ appears in Eq.(3.6). 
The difference between these curves within the 
droplet is the solvation contribution. 
\vskip 4mm

\noindent
FIG. 9.  Normalized free energy $\beta W_c$ without $v_{\rm is}$  
at $T/T_c=0.928$ (left) 
and $0.675$ (right) 
as a function of $A$. 
The points ($\bullet$)  (KO)  are 
obtained by solving  Eq.(2.21). 
The solid line (KO') is calculated from 
Eq.(3.14), while the broken 
line (Thomson) is 
from the first line of  Eq.(3.18).
\vskip 4mm

\noindent
FIG. 10.  Normalized free energy $\beta W_c$  
   with $v_{\rm is}$  
$T/T_c=0.928$ (left) 
and $0.675$ (right) 
as a function of $A$. 
The points ($\bullet$)  (KO)  are 
obtained by solving Eq.(2.21). 
The solid line (KO') is calculated from 
Eq.(3.14), while the broken 
line (Thomson) is 
from the first line of  Eq.(3.20).
\vskip 4mm

\clearpage

\begin{figure}[t]
\vspace*{5cm}
\epsfxsize=3.4in 
\centerline{\epsfbox{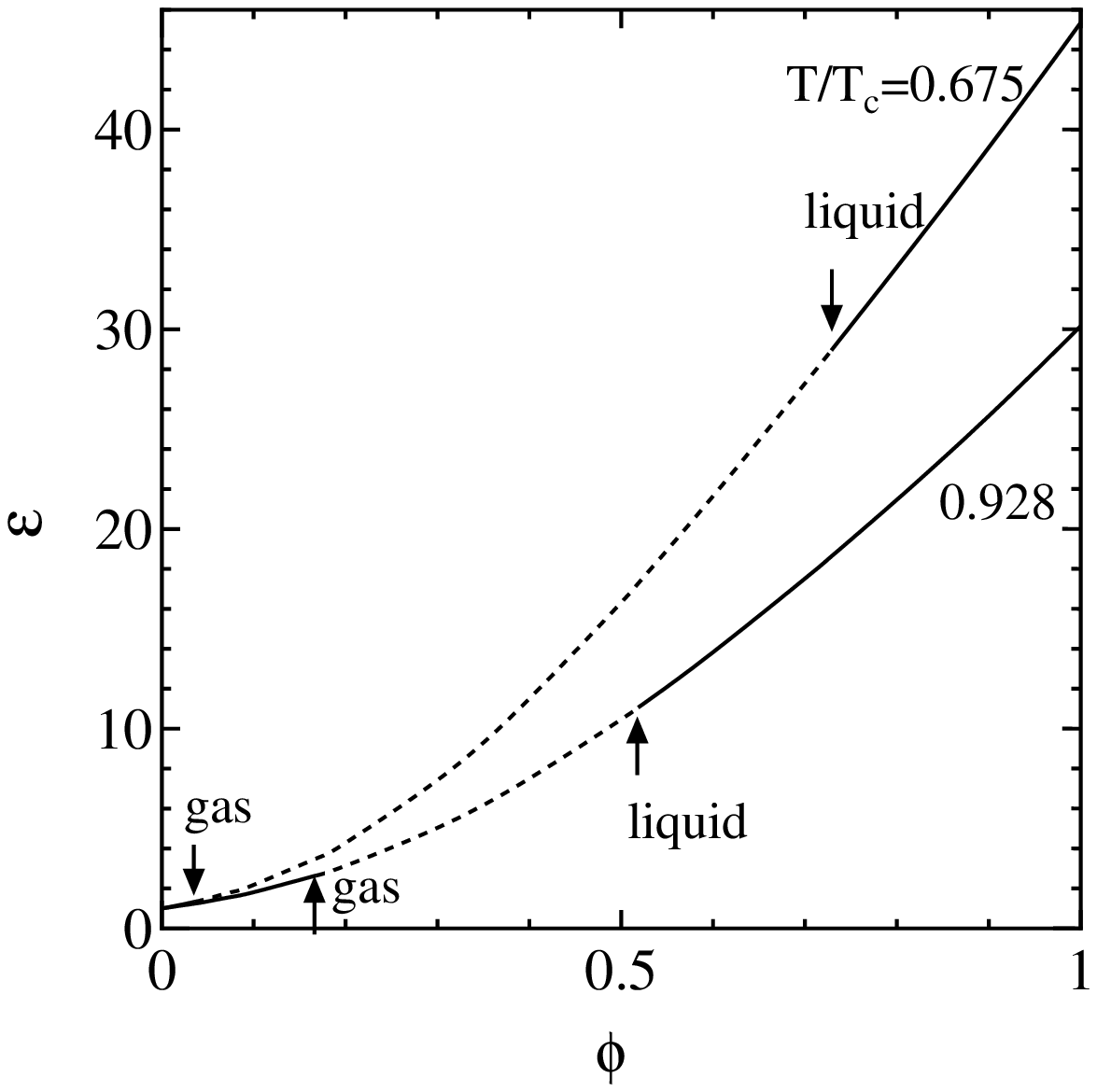}}
\caption{\protect
}
\label{1}
\end{figure}
\clearpage

\begin{figure}[t]
\vspace*{5cm}
\epsfxsize=4in 
\centerline{\epsfbox{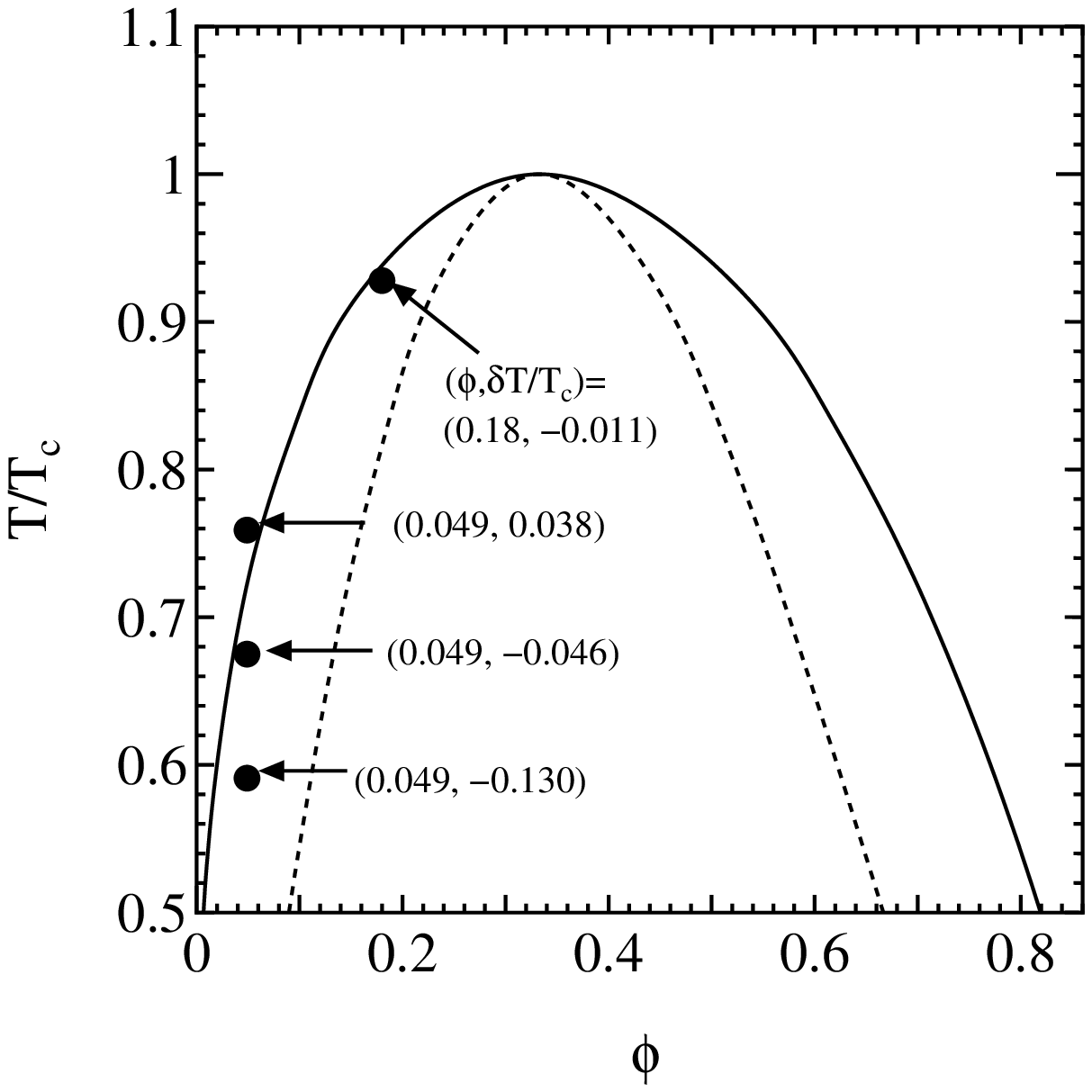}}
\caption{\protect
}
\label{2}
\end{figure}
\clearpage

\begin{figure}[t]
\vspace*{5cm}
\epsfxsize=3.4in 
\centerline{\epsfbox{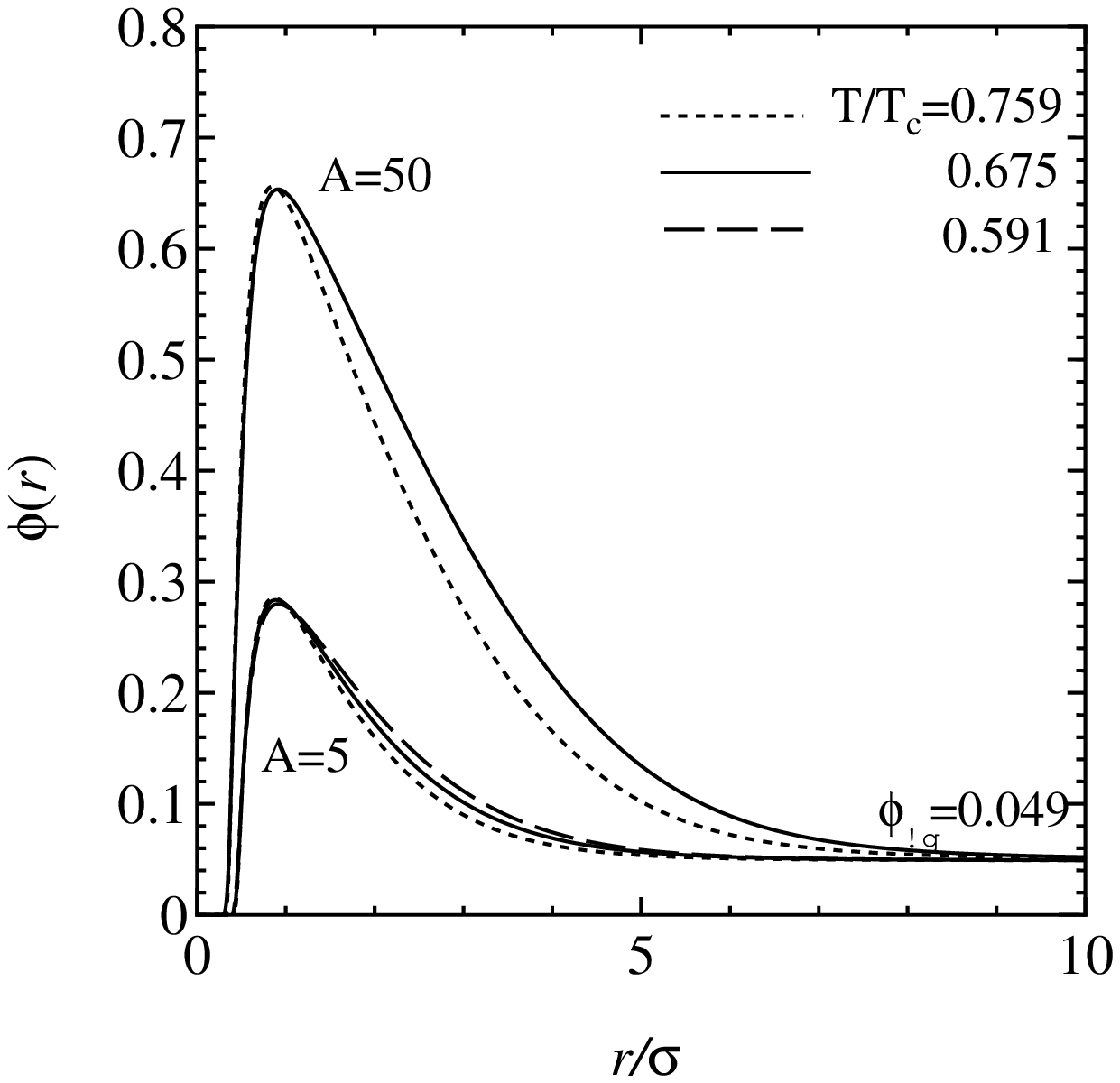}}
\caption{\protect
}
\label{3}
\end{figure}
\clearpage

\begin{figure}[t]
\vspace*{5cm}
\epsfxsize=6.2in 
\centerline{\epsfbox{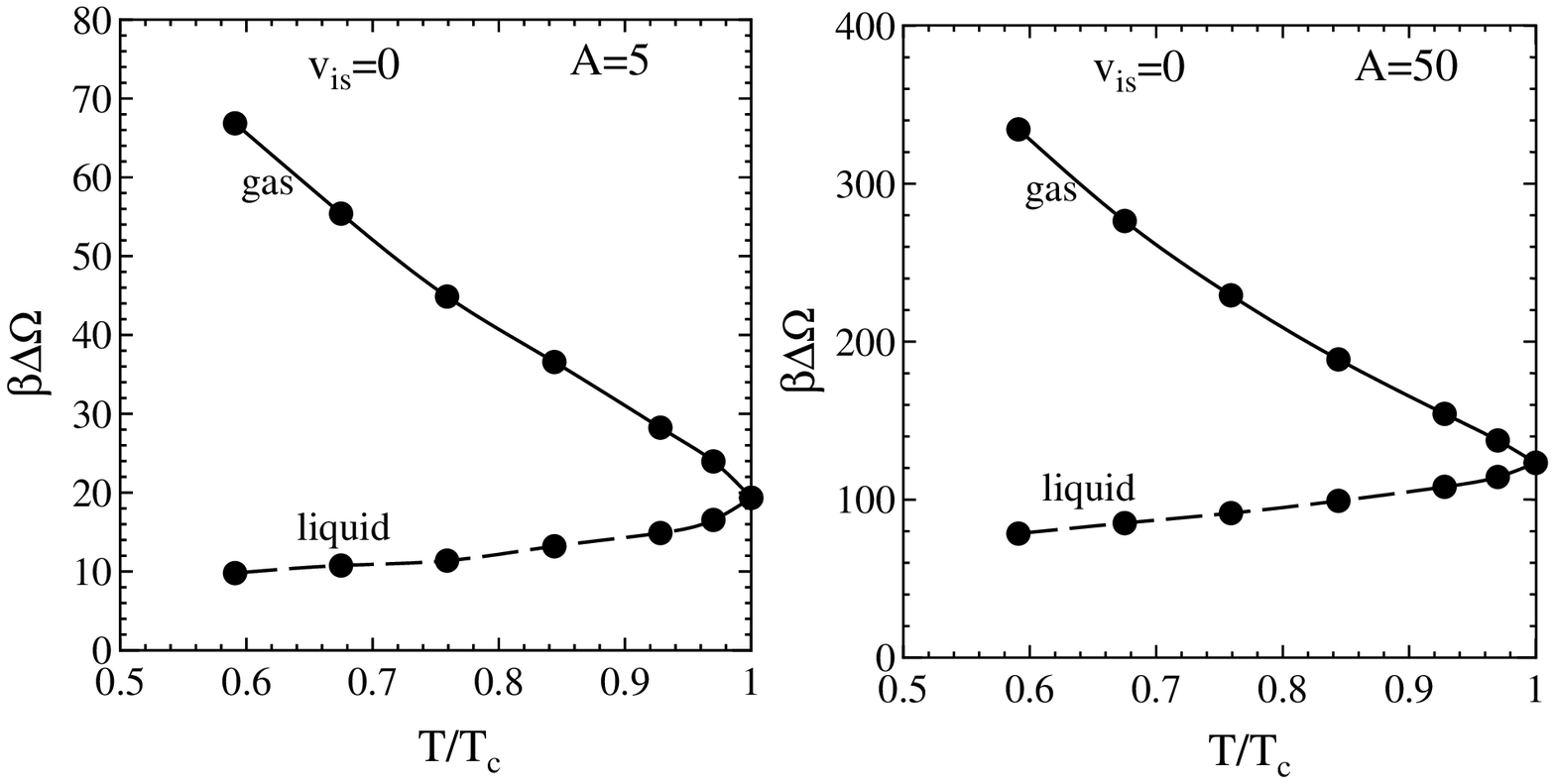}}
\caption{\protect
}
\label{5}
\end{figure}
\clearpage

\begin{figure}[t]
\vspace*{5cm}
\epsfxsize=6.2in 
\centerline{\epsfbox{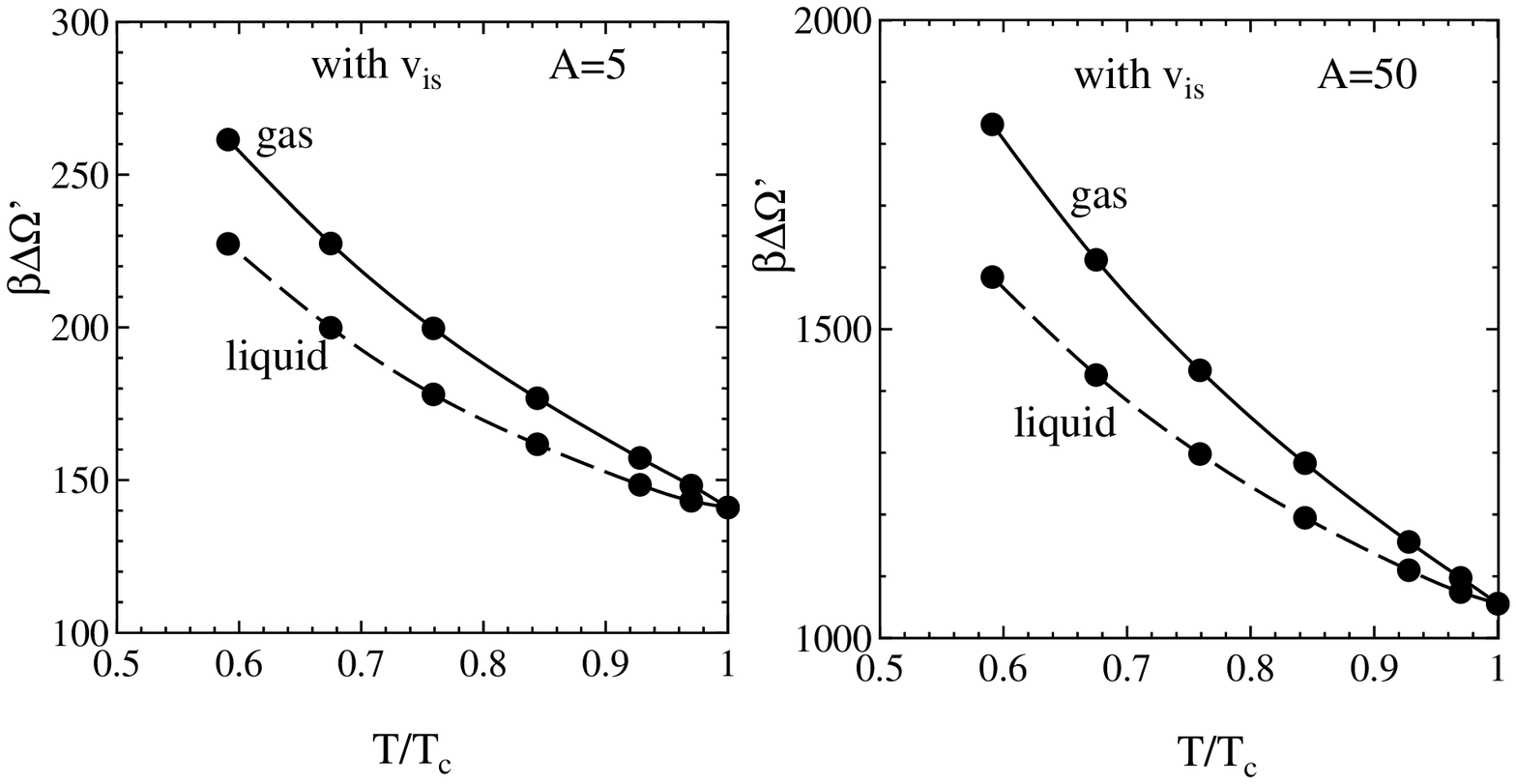}}
\caption{\protect
}
\label{6}
\end{figure}
\clearpage

\begin{figure}[t]
\vspace*{5cm}
\epsfxsize=6.2in 
\centerline{\epsfbox{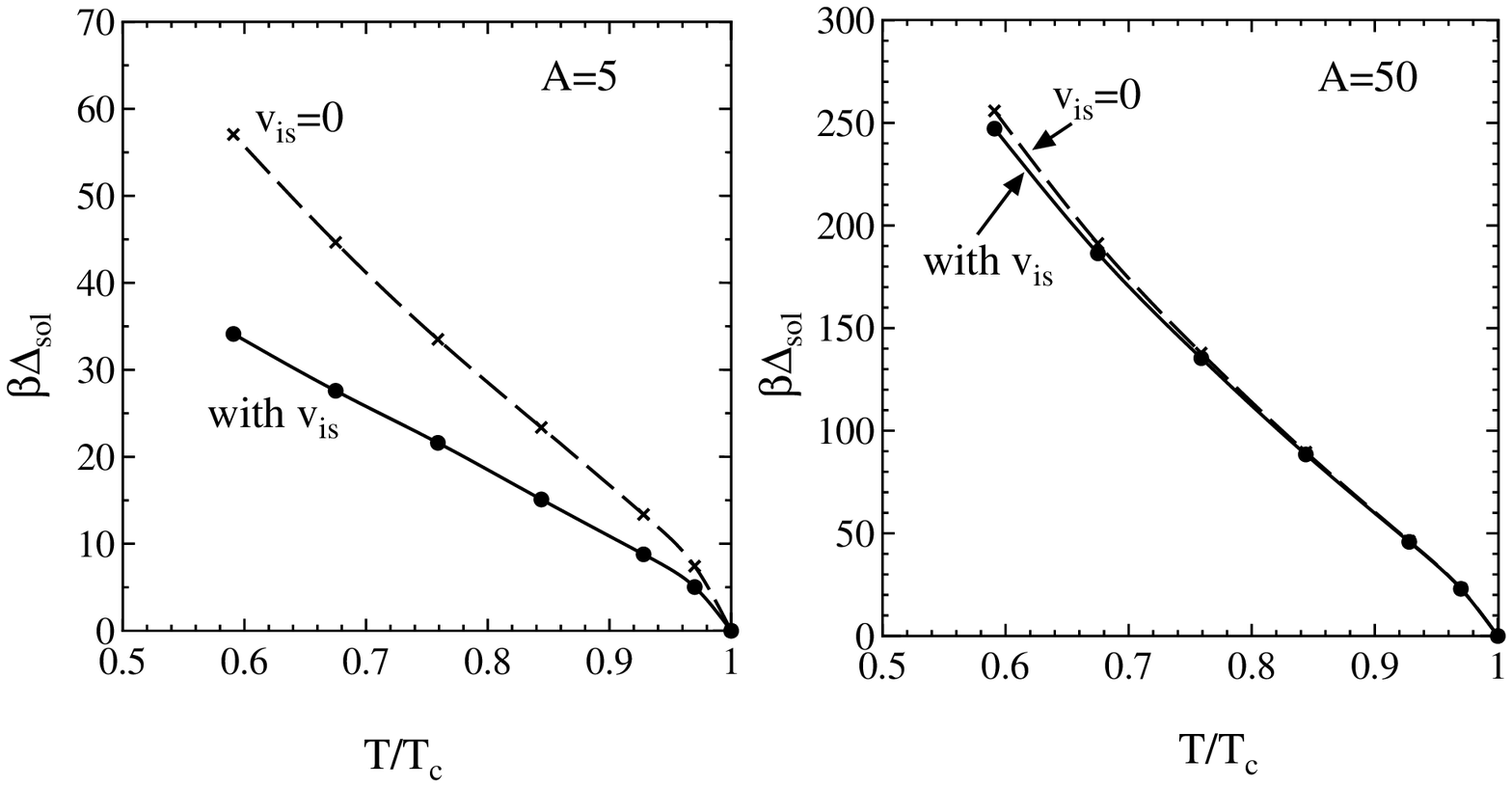}}
\caption{\protect
}
\label{7}
\end{figure}
\clearpage

\begin{figure}[t]
\vspace*{5cm}
\epsfxsize=6.2in 
\centerline{\epsfbox{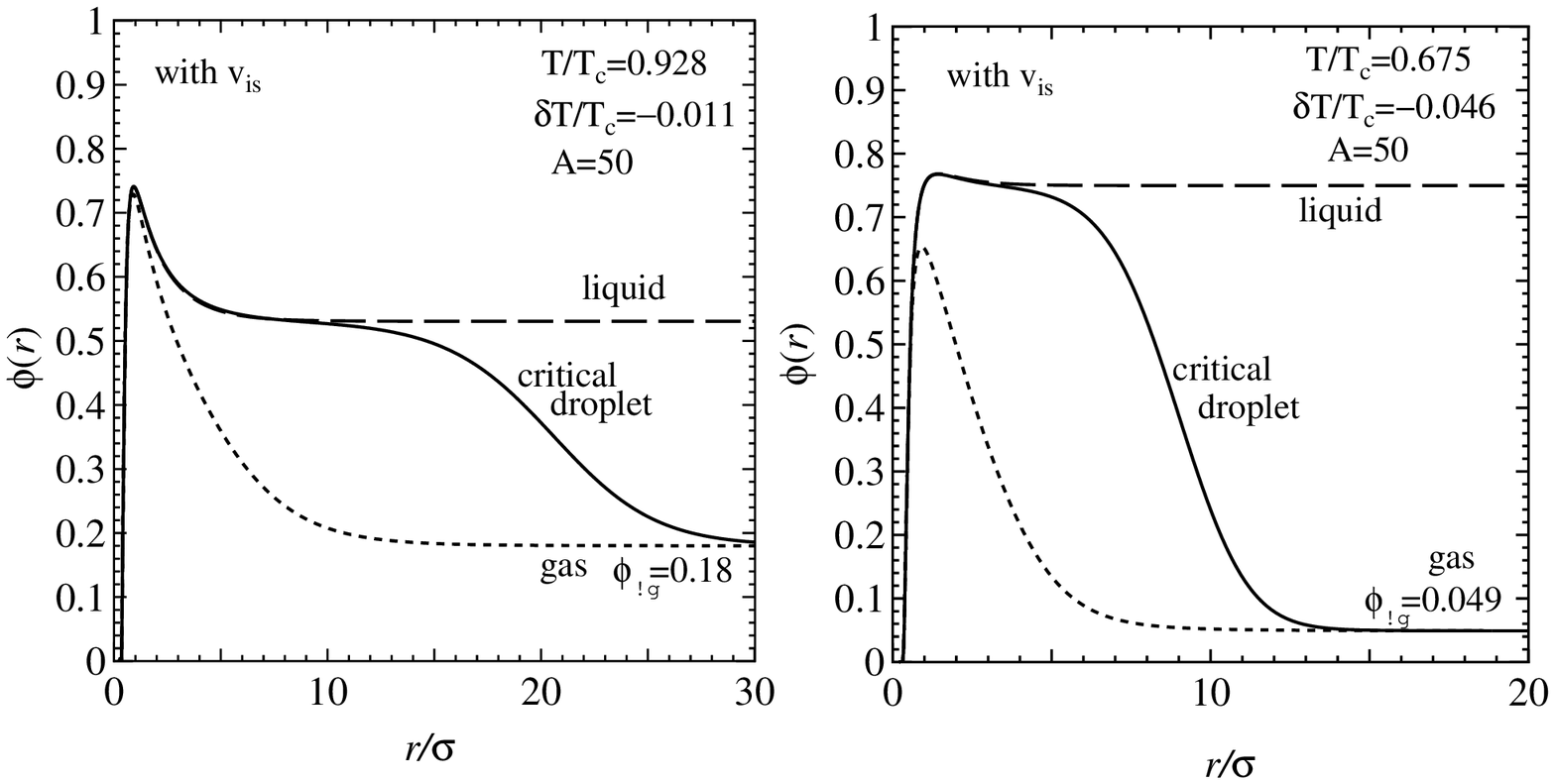}}
\caption{\protect
}
\label{4}
\end{figure}
\clearpage

\begin{figure}[t]
\vspace*{5cm}
\epsfxsize=3.2in 
\centerline{\epsfbox{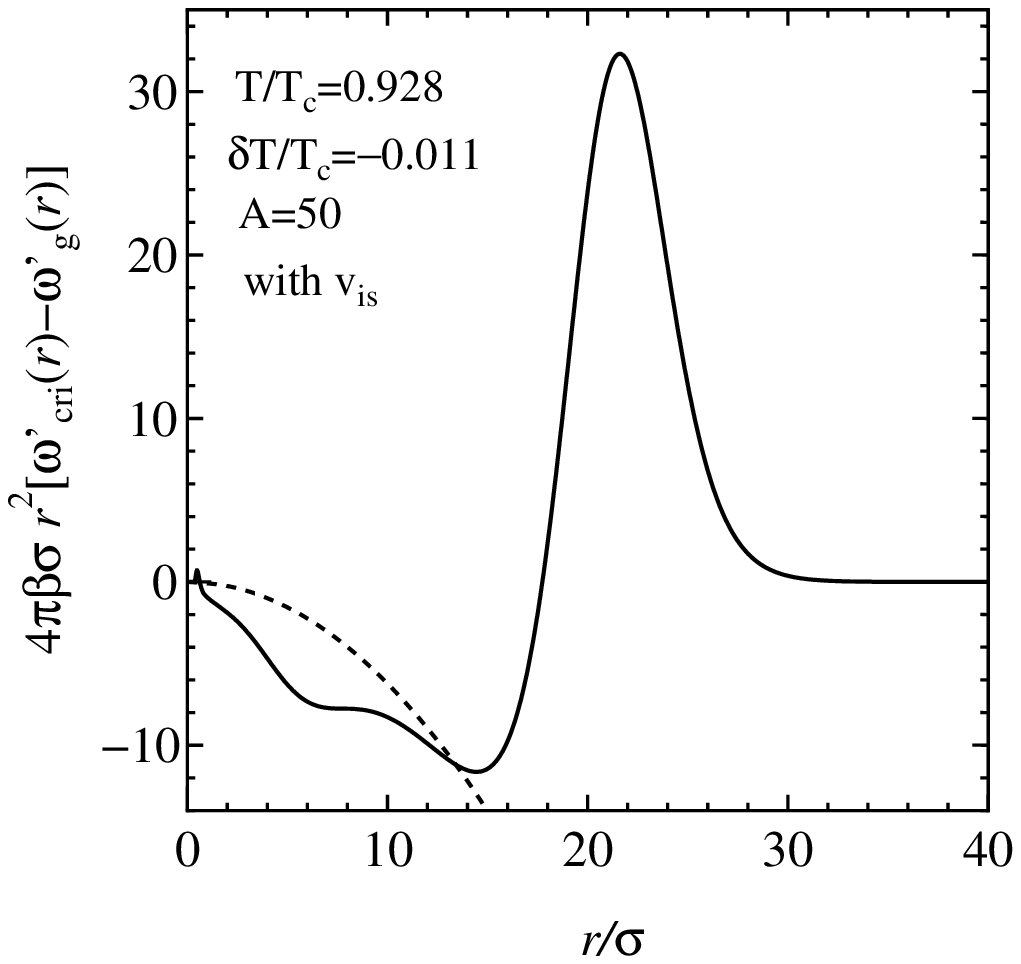}}
\caption{\protect
}
\label{4}
\end{figure}
\clearpage

\begin{figure}[t]
\vspace*{5cm}
\epsfxsize=6.2in 
\centerline{\epsfbox{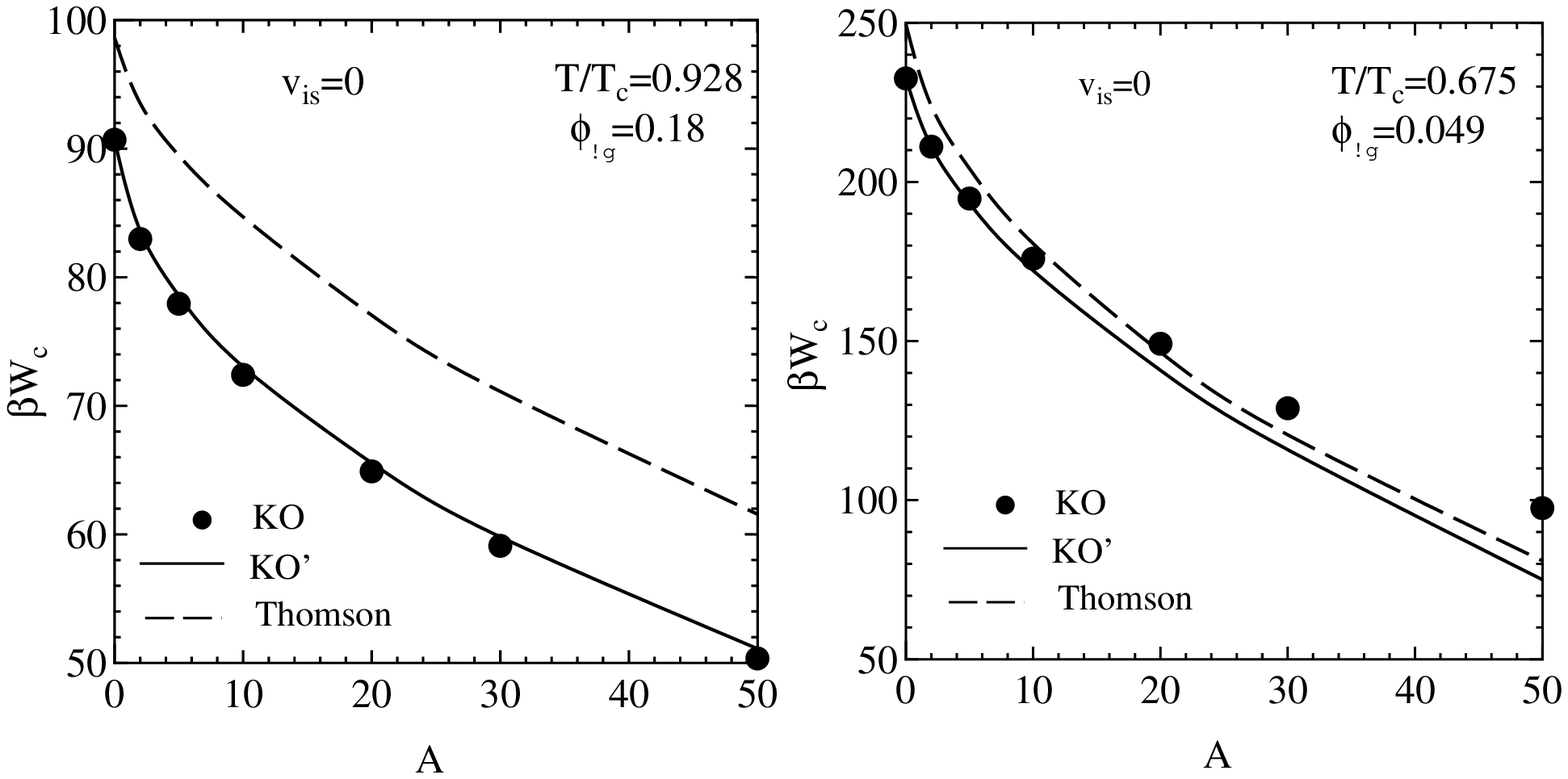}}
\caption{\protect
}
\label{8}
\end{figure}
\clearpage

\begin{figure}[t]
\vspace*{5cm}
\epsfxsize=6.2in 
\centerline{\epsfbox{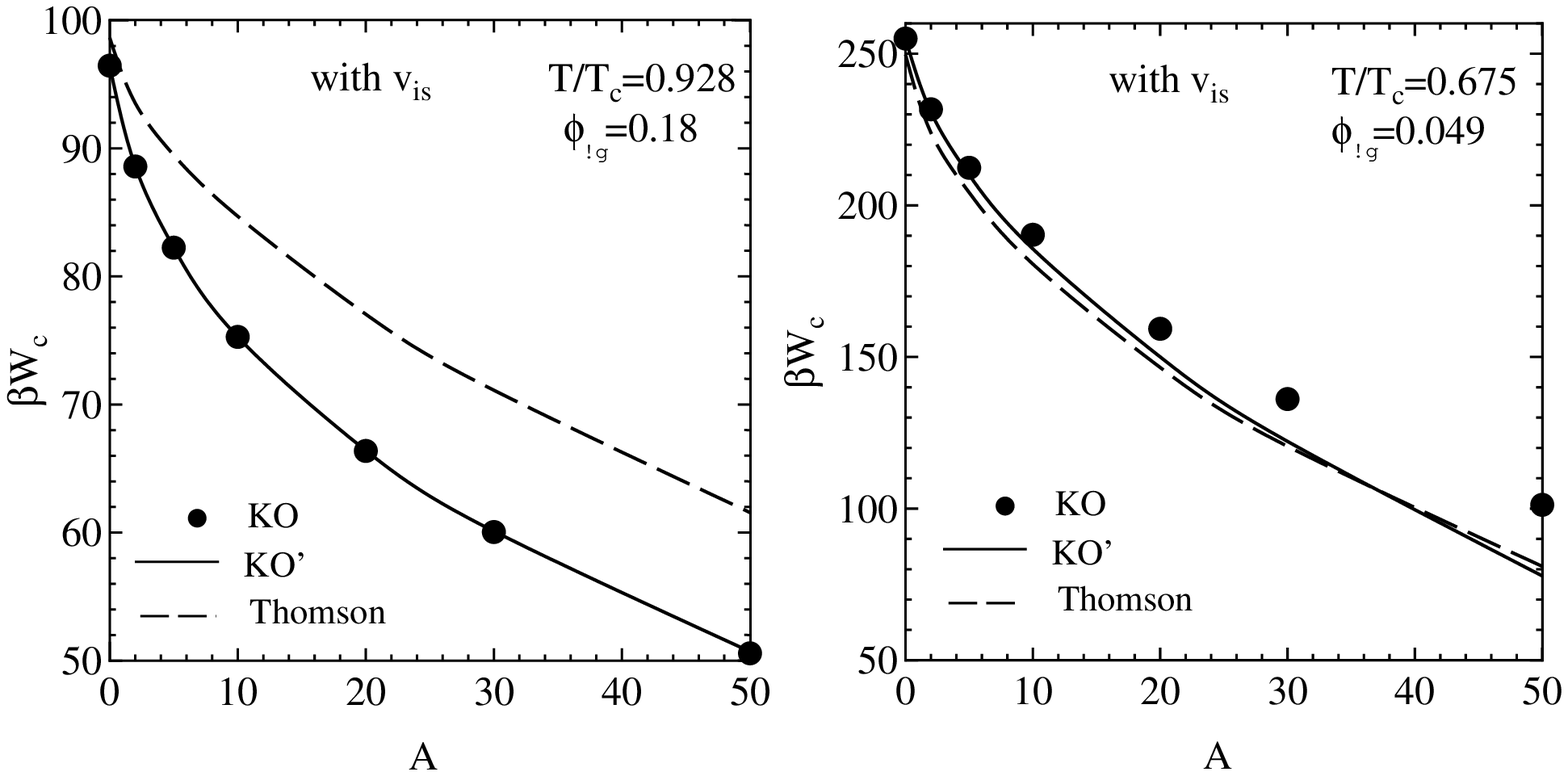}}
\caption{\protect
 }
\label{9}
\end{figure}
\clearpage


\begin{references}
\bibitem{comment} 
The  free energy 
due to the polarization and the charge \cite{NATO}  
is $F_e= \int d{\bi r}(
{\bi P}^2/2\chi+ {\bi E}^2/8\pi)$ 
with $\nabla\cdot({\bi E}+4\pi{\bi P})=4\pi\rho$. 
For ${\bi P}= \chi {\bi E}$ we obtain Eqs.(2.1) and (2.5). 
If ${\bi P}={\bi 0}$, 
it is equal to the vacuum result $
Z^2e^2/2R_{\rm B}$ for a single ion.  
If  ${\bi P}=\chi{\bi E}$ without  electrostriction, 
it is equal to  $\Delta\Omega_{{\rm Born}}$ in Eq.(2.17).
In the literature \cite{Is,Born} the 
difference  of these two quantities 
is  the Born formula for the solvation free energy.
\bibitem{however} In discussing the droplet growth 
we need to consider inhomogeneity of 
the temperature induced by latent heat absorption 
or generation at the interface \cite{Onukibook}.
\bibitem{cutoff} In  Eq.(2.17) 
we obtain  $R_{\rm B}=
(5/6) R_{\rm i}$  from Eq.(2.5) if 
the integration is performed also 
in the  region $r<R_{\rm i}$\cite{OK}.
\bibitem{grow} 
In dynamics 
 growing or shrinking of the droplet occurs depending 
on whether the added particle number is positive or negative.  
For example, 
this can be verified if we assume the time-evolution 
equation $\p n/\p t= \nabla^2 \delta (\Delta\Omega')/\delta n$.
\bibitem{critical} 
Near the critical point 
we obtain  $R_{\rm R}/\xi \sim  (1-T/T_c)^{(\nu +\beta)/3} 
$, where $\xi \propto  
(1-T/T_c)^{-\nu}$ is the correlation 
length with 
$\nu\cong 0.63$ and $\beta\cong 0.33$ being  the critical 
exponents \cite{Onukibook}. 
The interface thickness is of order $\xi$.
\end{references}
\end{document}